\documentclass[12pt]{iopart}
\usepackage{amssymb,graphicx}

\newcommand{\dd}{\mathrm{d}}
\newcommand{\pd}[2]{\frac{\partial #1}{\partial #2}}

\newcommand{\mean}[1]{\langle #1 \rangle}
\newcommand{\Int}[1]{\int\dd #1\;}
\newcommand{\IInt}[3]{\int_{#2}^{#3}\dd #1\;}
\newcommand{\Path}[1]{\int[\dd #1]\;}

\newcommand{\eps}{\varepsilon}
\newcommand{\lam}{\lambda}
\newcommand{\gam}{\gamma}
\newcommand{\kap}{\kappa}
\newcommand{\X}{\Gamma}
\newcommand{\sm}{\hat L}
\newcommand{\smm}{\hat L_\mathrm{m}}
\newcommand{\sms}{\hat L_\mathrm{s}}
\newcommand{\smed}{\Delta s_\mathrm{m}}
\newcommand{\st}{\Delta s_\mathrm{tot}}

\newcommand{\peq}{p_\mathrm{eq}}
\newcommand{\ps}{p_\mathrm{s}}
\newcommand{\As}{A_\mathrm{s}}
\newcommand{\Aa}{A_\mathrm{a}}

\begin{document}


\title[Stochastic Thermodynamics for Non-Markovian Processes]{Jarzynski
  Relation, Fluctuation Theorems, and Stochastic Thermodynamics for
  Non-Markovian Processes}

\author{T Speck and U Seifert}
\address{II. Institut f\"ur Theoretische Physik, Universit\"at
  Stuttgart, Pfaffenwaldring 57, 70550 Stuttgart, Germany}

\begin{abstract}
  We prove the Jarzynski relation for general stochastic processes including
  non-Markovian systems with memory. The only requirement for our proof is the
  existence of a stationary state, therefore excluding non-ergodic systems. We
  then show how the concepts of stochastic thermodynamics can be used to prove
  further exact non-equilibrium relations like the Crooks relation and the
  fluctuation theorem on entropy production for non-Markovian dynamics.
\end{abstract}

\maketitle


\section{Introduction}

The Jarzynski relation~\cite{jarz97} connects non-equilibrium work values $W$
spent in driving a system initially in equilibrium with the change of free
energy $\Delta F\equiv F_B-F_A$ between initial ($A$) and final state ($B$)
through the nonlinear average
\begin{equation}
  \label{eq:jr}
  \mean{e^{-\beta W}} = e^{-\beta\Delta F}.
\end{equation}
Here, $\beta\equiv(k_\mathrm{B}T)^{-1}$ is the inverse temperature of the heat
bath the system is coupled to and $k_\mathrm{B}$ is Boltzmann's constant. The
Jarzynski relation has found wide-spread application in both experiments and
computer simulations (for reviews, see~\cite{bust05,rito06}).
While~(\ref{eq:jr}) was derived originally for deterministic
dynamics~\cite{jarz97} and then generalized to stochastic Markov
processes~\cite{jarz97a}, its extension to general non-Markovian dynamics is
still open. The special case of a driven harmonic oscillator has been treated
analytically and numerically in~\cite{mai07}. In~\cite{zamp05}, the case of
non-equilibrium baths with memory is discussed for the entropy production.

In this paper, we will first give a general proof of the Jarzynski relation
which basically extends a previous proof for Markovian dynamics~\cite{impa05}.
We then specialize to Gaussian noise and discuss the role of time reversal.
This will allow us to show that further non-equilibrium relations like the
Crooks relation~\cite{croo99,croo00} and the detailed fluctuation theorem in
non-equilibrium steady states~\cite{kurc98,lebo99,seif05a} hold as well.
Together with their counterparts holding for deterministic thermostated
dynamics~\cite{evan93,gall95,evan02}, all these non-equilibrium relations show
a surprising robustness against the underlying dynamics.


\section{Proof of the Jarzynski relation for general stochastic processes}

Let the energy of the system be given by a Hamiltonian $H(\X,\lam)$, where
$\X$ is a point in phase space.  We assume that we can control the system
externally through a change of the parameter $\lam$ where the function
$\lam(\tau)$ is called the protocol. We will consider trajectories $\X(\tau)$
in the interval
\begin{equation}
  t_0 \leqslant 0 \leqslant \tau \leqslant t \leqslant t_1
\end{equation}
involving four times. At the lower boundary $t_0$ we prepare the system such
that it retains no memory of earlier times. We then observe single
trajectories from $t_0$ to $t_1$. However, we will drive the system through a
change of $\lam$ only during the inner interval $0\leqslant\tau\leqslant t$
such that $\lam$ is constant outside. The change of energy identified as the
work spent along a single trajectory $\X(\tau)$ is then
\begin{equation}
  \label{eq:work}
  W[\X(\tau);\lam(\tau)] \equiv 
  \IInt{\tau}{0}{t}\dot\lam(\tau)\pd{H}{\lam}(\X(\tau),\lam(\tau)).
\end{equation}
Hence, the work depends only on the inner section of the trajectory, which is
the first ingredient for the proof. In the following, we will drop the
implicit dependence on the protocol in the argument of functionals.

The second ingredient for the proof is the time evolution equation
\begin{equation}
  \label{eq:evo}
  \partial_\tau p(\X,\tau) = \sm(\tau;t_0)p(\X,\tau)
\end{equation}
of the distribution $p(\X,\tau)$ determining the probability to find the
system in a specific region of phase space. The evolution of this distribution
is governed by the operator $\sm(\tau;t_0)$. In the Markov case, the operator
$\sm(\tau;t_0)\rightarrow\smm(\lam(\tau))$ is the generator of a
semi-group~\cite{hang82}. It then completely defines the stochastic process.
It is independent of $t_0$ and therefore it does not depend on the details of
preparation while it depends on time due to the change of the external
parameter $\lam$.

It is somewhat surprising that the same, apparently time-local,
equation~(\ref{eq:evo}) holds also for non-Markovian
processes~\cite{hang77,hang82}. This can be understood by realizing that the
complete information about processes with memory is contained in the
transition probability depending on the whole history rather than in the
single-point distribution $p(\X,\tau)$. We denote with $\hat
U(\tau'|\tau;t_0)$ the operator that propagates the system from time
$\tau<\tau'$ to the later time $\tau'$. The propagator actually depends on the
whole function $\lam(\tau)$ up to $\tau'$ since any change of the protocol
will have consequences for the following evolution. From the propagator, we
can define the operator
\begin{equation}
  \label{eq:sub}
  \sms(\tau;t_0) \equiv \partial_{\tau'}
  \left. \hat U(\tau'|\tau;t_0)\right|_{\tau'=\tau^+}
\end{equation}
describing a ``substitute'', non-stationary Markov process,
$\sm(\tau;t_0)\rightarrow\sms(\tau;t_0)$, which leads to the same single-point
distribution $p(\X,\tau)$ but to a different transition probability than the
non-Markovian process~\cite{hang77}. In particular, knowledge of the
operator~(\ref{eq:sub}) is not sufficient to calculate correlation functions.
In contrast to the Markov case, the dependence on the control parameter $\lam$
of the operator~(\ref{eq:sub}) is implicit. In the appendix, we give an
explicit example for such a substitute operator.

We restrict our proof to dynamics with a unique steady state, i.e., for fixed
$\lam$ the system will relax towards a unique probability distribution
$\ps(\X,\lam)$ depending on the control parameter,
$\lim_{\tau\rightarrow\infty}p(\X,\tau)\rightarrow\ps(\X,\lam)$. This is
equivalent to ergodic processes with or without memory (see Ref.~\cite{bao05}
for a discussion of non-Markovian processes which break ergodicity). In the
absence of non-conservative driving, the stationary distribution must be the
equilibrium Gibbs-Boltzmann distribution
\begin{equation}
  \label{eq:eq}
  \peq(\X,\lam) = [Z(\lam)]^{-1} e^{-\beta H(\X,\lam)},
\end{equation}
where the partition function
\begin{equation}
  Z(\lam) = \Int{\X}e^{-\beta H(\X,\lam)}
\end{equation}
determines the free energy $F(\lam)=-\beta^{-1}\ln Z(\lam)$. Then $Z_{A,B}$ is
the partition function of the initial and the final state, respectively.

The third ingredient to the proof is the property that the equilibrium
distribution is the stationary solution
\begin{equation}
  \label{eq:erg}
  \sm(\tau;t_0)\peq(\X,\lam(\tau)) = 0
\end{equation}
for the corresponding value $\lam=\lam(\tau)$ of the control parameter.
Whereas this is evident in the case of a Markovian operator, due to the
implicit dependence on $\lam$ it is not so obvious in the non-Markovian case
and we give a proof by contradiction. First we note that for a proper
Markovian substitute process, the operator~(\ref{eq:sub}) must have a
stationary solution. Now suppose that at time $\tau'$ we stop the process and
hold the parameter fixed with value $\lam=\lam(\tau')$. Under very general
conditions, which are fulfilled by any transition probability, the
Perron-Frobenius theorem ensures that the propagator $\hat U(\tau|\tau';t_0)$
has an eigenstate $p_1(\X;\tau,\tau')$ corresponding to the eigenvalue $1$
depending on $\tau'$ and in principle also depending on $\tau$, i.e.,
\begin{equation}
  \label{eq:eigen}
  \hat U(\tau|\tau';t_0)p_1(\tau,\tau') = p_1(\tau,\tau').
\end{equation}
Furthermore, this eigenstate $p_1(\X;\tau,\tau')$ is a normalized,
non-negative probability distribution. From the definition~(\ref{eq:sub}), we
calculate
\begin{equation}
  \label{eq:null}
  \sms(\tau';t_0) p_1(\tau,\tau') = \lim_{\eps\rightarrow 0}\frac{1}{\eps}
  \left[\hat U(\tau'+\eps|\tau';t_0)p_1(\tau,\tau')-p_1(\tau,\tau')\right] 
  \neq 0
\end{equation}
which is non-zero for both arbitrary functions $p_1$ and for the eigenfunction
$p_1(\tau,\tau')$ of the propagator if the latter depends on $\tau$ since
$\tau$ does not match the leading time argument of the propagator. This would
mean that the substitute operator~(\ref{eq:sub}) has no stationary solution.
This contradiction is resolved only if the eigenfunction $p_1(\tau')$ is
independent of $\tau$. Moreover, taking then the limit $\tau\rightarrow\infty$
in~(\ref{eq:eigen}), we find from the ergodicity condition that
$p_1(\X,\tau')=\peq(\X,\lam)$. Finally, we note that due to causality, we do
not have to actually stop the process at a $\tau'$ since the system cannot
depend on the future protocol and~(\ref{eq:erg}) must hold for all times
$\tau$.

With these three ingredients, the proof of the Jarzynski
relation~(\ref{eq:jr}) becomes simple. We prepare the system at time
$t_0\leqslant 0$ in equilibrium and start to drive the system at $\tau=0$
until $\tau=t$ following the protocol $\lam(\tau)$.  Inspecting the expression
for the work~(\ref{eq:work}), we see that its instantaneous change $\dot
H(\X,\tau)\equiv\dot\lam\partial_\lam H(\X,\lam)$ only depends on the actual
state $\X$ the system is in. Hence, the operator $\sm(\tau;t_0)$ is all we
need to prove the Jarzynski relation including non-Markovian processes. To
this end, we consider the joint probability $\rho(\X,w,\tau)$ for finding the
system in state $\X$ at time $\tau$ and for having accumulated an amount of
work $w$ up to this time~\cite{impa05,spec04}.  A change of the state or the
control parameter $\lam$ will lead to a probability current $j_w=\dot H\rho$
in the direction of the $w$ coordinate, hence the equation of motion for the
joint probability becomes
\begin{equation}
  \partial_\tau\rho(\X,w,\tau) = 
  \left[\sm(\tau;t_0)-\dot H(\X,\tau)\partial_w\right]\rho(\X,w,\tau)
\end{equation}
due to the conservation of probability. We can prove the Jarzynski
relation~(\ref{eq:jr}) by first defining the function
\begin{equation}
  \psi(\X,\tau) \equiv \IInt{w}{-\infty}{+\infty} \rho(\X,w,\tau)e^{-\beta w}.
\end{equation}
Since the probability for extreme work values $w\rightarrow\pm\infty$
vanishes, after one integration by parts the equation of motion becomes
\begin{equation}
  \partial_\tau\psi(\X,\tau) = \left[\sm(\tau;t_0) 
    - \beta\dot H(\X,\tau)\right]\psi(\X,\tau).
\end{equation}
A solution of this equation is the equilibrium Boltzmann factor
\begin{equation}
  \label{eq:sol}
  \psi(\X,\tau) = Z_A^{-1}e^{-\beta H(\X,\lam(\tau))}
\end{equation}
provided $\sm(\tau;t_0)\psi(\X,\tau)=0$ holds for all $\tau$ as discussed
above, see~(\ref{eq:erg}). The solution~(\ref{eq:sol}) obeys both the initial
equilibrium condition $\psi(\X,0)=\peq(\X,\lam(0))$ and
$\Int{\X}\psi(\X,t)=Z_B/Z_A$ which implies the Jarzynski
relation~(\ref{eq:jr}).

We have thus proved that the Jarzynski relation~(\ref{eq:jr}) holds for any
kind of non-Markovian noise by exploiting the existence of a time-local
substitute operator~(\ref{eq:evo}) which annihilates the Gibbs-Boltzmann
distribution $\peq(\X,\lam)$ for any $\lam$ reached during the process. This
proof only involves the work definition~(\ref{eq:work}) but no other
thermodynamic notions like heat and entropy, which we will discuss now.


\section{Stochastic thermodynamics}

The Jarzynski relation~(\ref{eq:jr}) can be embedded into the larger framework
of stochastic thermodynamics.  The crucial idea is to extent the notion of
work and heat to small, stochastic systems coupled to a heat bath in the
following way: one first identifies the energy change caused externally as the
work and then the heat dissipated due to the interaction with the bath follows
from the first law~\cite{seki98}. The first law thus reads
\begin{equation}
  \label{eq:fl}
  Q[\X(\tau)] \equiv W[\X(\tau)] - \Delta H
\end{equation}
with the change of internal energy $\Delta H\equiv
H(\X(t_1),\lam(t))-H(\X(t_0),\lam(0))$ along the specific trajectory
$\X(\tau)$. The sign of the heat is convention, here we choose it to be
positive if energy is dissipated into the bath. Using the definition of the
work~(\ref{eq:work}), we write the right hand side of equation~(\ref{eq:fl})
under one integral sign. By inserting the total derivative of the energy, the
heat becomes the functional
\begin{equation}
  \label{eq:heat}
  Q[\X(\tau)] =
  - \IInt{\tau}{t_0}{t_1}\dot\X(\tau)\cdot\pd{H}{\X}(\X(\tau),\lam(\tau)).
\end{equation}
So far, we did not make any assumptions about the bath and the dynamics of the
system and hence the expressions for work and heat should hold for both
Markovian and non-Markovian dynamics.


\section{Time reversal}

The identification of the heat allows for a second route to derive the
Jarzynski relation~(\ref{eq:jr}) via time
reversal~\cite{croo00,seif05a,maes03}. For notational simplicity, we restrict
our discussion to one overdamped degree of freedom $\X=x$ moving in the
potential $H=V(x,\lam)$. We define the functional
\begin{equation}
  \label{eq:R}
  R[x(\tau)] \equiv \ln\frac{P[x(\tau)]}{P[\tilde x(\tau)]} =
  \ln\frac{P[x(\tau)|x_0]\peq(x_0)}{P[\tilde x(\tau)|x_1]\peq(x_1)}
\end{equation}
which fulfills the relation $\mean{\exp[-R]}=1$ by definition since
$P[x(\tau)]$ is the probability of a trajectory $x(\tau)$. In the second step,
we have separated the initial state $x_0\equiv x(t_0)$ from the conditional
probability $P[x(\tau)|x_0]$ times the probability distribution of the initial
state. The path $\tilde x(\tau)\equiv x(t_0+t_1-\tau)$ denotes the time
reversal starting in $x_1\equiv x(t_1)$.

We model the dynamics of a system coupled to a non-Markovian bath with
Gaussian noise $\eta$ through the generalized Langevin equation
\begin{equation}
  \label{eq:gle}
  \gam(t-\tau)\circ\dot x(\tau) = -V'(x(t),\lam(t)) + \eta(t)
\end{equation}
with friction kernel $\gam(\tau)$. The prime denotes derivation with respect
to $x$. For convenience, we define the operation
\begin{equation}
  g(\tau)\circ h(\tau) \equiv \IInt{\tau}{t_0}{t_1}g(\tau)h(\tau)
\end{equation}
where integration is carried out over the free variable $\tau$ analogous to
Einstein's sum convention. If one or both of the functions depend on $x$ then
by this short notation we mean $g(\tau)\equiv g(x(\tau),\lam(\tau))$.

Equation~(\ref{eq:gle}) includes the Markov case through choosing a time-local
friction kernel $\gam(\tau)=2\gam\delta(\tau)$ with friction coefficient
$\gam$. The bath correlation function is defined as
\begin{equation}
  \label{eq:C}
  C(\tau_1-\tau_2) \equiv \mean{\eta(\tau_1)\eta(\tau_2)} 
\end{equation}
and the system is guaranteed to equilibrate with the heat reservoir through
Kubo's second fluctuation-dissipation theorem~\cite{kubo}
\begin{equation}
  \label{eq:kubo}
  \gam(\tau) = \cases{\beta C(\tau) & $\tau \geqslant 0$, \cr 
    0 & $\tau < 0$.}
\end{equation}
The probability of a certain noise history $\eta(\tau)$ obeys
$P[\eta(\tau)]>0$ for any continuous path $\eta(\tau)$ and normalization
$\Path{\eta(\tau)} P[\eta(\tau)]=1$ with functional measure $[\dd\eta(\tau)]$.
Gaussian noise is completely defined by its (zero) mean and correlations
$C(\tau)$. We therefore write $P[\eta(\tau)]=\exp\{-A[\eta(\tau)]\}$
introducing the quadratic ``action'' functional
\begin{equation}
  A[\eta(\tau)] = \frac{1}{2} 
  \eta(\tau_1)\circ K(\tau_1-\tau_2)\circ\eta(\tau_2).
\end{equation}
The symmetric noise kernel $K(\tau)=K(-\tau)$ is the operator inverse of the
bath correlation function $C(\tau)$,
\begin{equation}
  \label{eq:inv}
  C(\tau_1-\tau) \circ K(\tau-\tau_2) = \delta(\tau_1-\tau_2).
\end{equation}
We change variables from $\eta$ to $x$ with the probability of a single
trajectory
\begin{equation}
  P[x(\tau)|x_0] = J[x(\tau)]\exp{\{-\As[x(\tau)]-\Aa[x(\tau)]\}}.
\end{equation}
This change of variables makes it necessary to consider the conditional
probability since the noise history does not determine the initial state
$x_0$. The total action becomes a sum of two terms defined as
\begin{equation}
  \fl
  \As[x(\tau)] \equiv
  \frac{1}{2} V'(\tau_1)\circ K(\tau_1-\tau_2)\circ V'(\tau_2) + 
  \left[ \gam(\tau_1-\tau)\circ\dot x(\tau) \right] 
  \circ K(\tau_1-\tau_2) \circ \left[ \gam(\tau_2-\tau)\circ\dot x(\tau) 
  \right]
\end{equation}
and
\begin{equation}
  \label{eq:Aa}
  \Aa[x(\tau)] \equiv \left[ \gam(\tau_1-\tau) \circ \dot x(\tau) \right]
  \circ K(\tau_1-\tau_2) \circ V'(\tau_2),
\end{equation}
where we have replaced the noise through the generalized Langevin
equation~(\ref{eq:gle}). This change of variables involves further the
Jacobian $J[x(\tau)]\equiv\det[\delta\eta(t)/\delta x(\tau)]$ given by the
functional determinant. For time reversal, we run along the trajectory
$x(\tau)$ in the opposite direction from $t_1$ to $t_0$. In the integrals,
this amounts both to the substitution $\dot x\mapsto-\dot x$ and to inverting
the time argument of the kernels. Under these operations, both the symmetric
action $\As$ and the Jacobian $J$ stay invariant. The antisymmetric action of
the time-reversed trajectory becomes
\begin{equation}
  \Aa[\tilde x(\tau)] = -\left[ \gam(\tau-\tau_1) \circ \dot x(\tau) \right]
  \circ K(\tau_1-\tau_2) \circ V'(\tau_2)
\end{equation}
and hence the sum is
\begin{equation}
  -\Aa[x(\tau)]+\Aa[\tilde x(\tau)] =
  -\dot x(\tau_1)\circ M(\tau_1-\tau_2) \circ V'(\tau_2)
\end{equation}
with kernel
\begin{equation}
  \label{eq:kern}
  M(\tau_1-\tau_2) = [\gam(\tau_1-\tau)+\gam(\tau_1-\tau)]\circ K(\tau-\tau_2).
\end{equation}
The sum in the square brackets equals the symmetric bath correlation function
$C(\tau_1-\tau)$ through use of the fluctuation-dissipation
theorem~(\ref{eq:kubo}). Following~(\ref{eq:inv}), the kernel then reduces to
the $\delta$-function. The functional $R$ finally reads
\begin{equation}
  \label{eq:RQ}
  \frac{1}{\beta} R[x(\tau)] 
  = Q[x(\tau)] + \Delta V - \Delta F = W[x(\tau)] - \Delta F
\end{equation}
with the heat~(\ref{eq:heat}) independent of the actual bath correlation
function $C(\tau)$. The only requirement is that the bath itself is and stays
in equilibrium as it is in the case of Markov processes.


\section{Discussion and the path to proving further non-equilibrium relations}

We can now discuss the role of the two times $t_0$ and $t_1$. In equilibrium,
fluctuations have no memory. Since the trajectory $\tilde x(\tau)$ is just the
mirror image of $x(\tau)$, their probabilities must then be equal,
$P[x(\tau)]=P[\tilde x(\tau)]$, with $R=1$. We can therefore chose an
arbitrary interval $t_0\leqslant\tau\leqslant t_1$ during which we observe the
trajectory and the heat fulfills $Q=-\Delta V$. When we now drive the system
through the manipulation of $\lam$, we can actually choose the driving
interval as the observation interval, $t_0\rightarrow0^-$ and $t_1\rightarrow
t^+$. However, the noise kernel $K(\tau)$ defined through~(\ref{eq:inv}) then
depends on the two times $t_0$ and $t_1$. This reflects the fact that both the
forward and the time-reversed trajectory are cut off although the force at the
boundary still remembers the velocity of earlier times.

Although we have shown the relation~(\ref{eq:RQ}) only for Gaussian noise, the
fact that in the first part of the paper we have proven the Jarzynski relation
without assuming a specific type of noise suggests that the
relation~(\ref{eq:RQ}) is also valid more generally. However, a direct
treatment of non-Gaussian noise within the path integral formalism seems
technically challenging.

The fact that~(\ref{eq:RQ}) holds also for non-Markovian dynamics implies the
validity of other non-equilibrium relations. First, from the
definition~(\ref{eq:R}), one can derive the Crooks relation~\cite{croo00}
\begin{equation}
  \label{eq:crooks}
  \frac{p_R(-W)}{p_F(+W)} = e^{-\beta(W-\Delta F)}
\end{equation}
if we distinguish between the probability distribution of the work
$p_{F,R}(W)$ spent in the forward and time-reversed processes, respectively.
Second, for an equilibrated bath, we can still identify the dissipated heat as
entropy change in the heat bath, $\smed=\beta Q$. With the change of entropy
of the system,
\begin{equation}
  \Delta s \equiv s(t_1) - s(t_0), \qquad
  s(\tau) \equiv -\ln p(\X(\tau),\tau),
\end{equation}
the fluctuation theorem for the total entropy production $\st=\smed+\Delta
s$~\cite{seif05a}
\begin{equation}
  \label{eq:seif}
  \mean{e^{-\st}} = 1
\end{equation}
remains valid.

Finally, our analysis of non-Markovian processes can be easily extended to
systems driven by nonconservative forces which for constant $\lam$ reach a
non-equilibrium steady state with probability distribution $\ps(\X)$. In this
case, we have to include the nonconservative forces $f$ in the external work,
leading to
\begin{equation}
  \label{eq:work:nc}
  W[\X(\tau)] \equiv 
  \IInt{\tau}{0}{t}\left[\dot\lam(\tau)\pd{H}{\lam}(\X(\tau),\lam(\tau))
    + f(\tau)\cdot\dot\X(\tau)
  \right].
\end{equation}
The heat is still determined through the first law~(\ref{eq:fl}). If we
generalize the functional $R$ from equation~(\ref{eq:R}) by replacing the
equilibrium distribution $\peq$ with the stationary distribution $\ps$, it is
straightforward to derive the relation
\begin{equation}
  \label{eq:RQ:nc}
  R[\X(\tau)] = \beta Q[\X(\tau)] + \ln\frac{\ps(\X(t_0))}{\ps(\X(t_1))},
\end{equation}
from which the integral fluctuation theorem for entropy
production~(\ref{eq:seif}) follows. Moreover, in the case of stationary
driving ($\dot\lam=0$) also the detailed fluctuation
theorem~\cite{kurc98,lebo99,seif05a}
\begin{equation}
  \label{eq:detailed}
  \frac{P(-\st)}{P(+\st)} = e^{-\st}
\end{equation}
follows for a finite time interval, where $P(\st)$ is the probability
distribution of the total entropy production. The relation~(\ref{eq:detailed})
is also found in deterministic steady state systems in the long-time
limit~\cite{evan93,gall95,evan02}.

\section{Summary and outlook}

As our main result, we have shown that the Jarzynski relation holds for
general ergodic systems governed by stochastic dynamics including
non-Markovian processes. We have further confirmed in the case of Gaussian
noise that the relations~(\ref{eq:RQ}) and~(\ref{eq:RQ:nc}) between the heat
$Q$ and the functional $R$, which serves as convenient starting point to
derive further exact non-equilibrium relations, still holds for non-Markovian
processes. Therefore this class of exact non-equilibrium relations, of which
the Jarzynski relation is arguably the most prominent, shows a surprising
robustness against the underlying dynamics.  An open question which will
require further investigation is to which extent the concepts discussed in
this paper can be generalized to non-ergodic systems.

\appendix
\section{Substitute operator for a moving trap}

As an illustration, we calculate the substitute operator in case of a particle
moving in one dimension with position $x$ which is trapped in a harmonic
potential $V(x,\lam)=(k/2)(x-\lam)^2$. The generalized Langevin
equation~(\ref{eq:gle}) then becomes linear and can be solved by Laplace
transformation as
\begin{equation*}
  x(t) = G_1(t)x_0 + \IInt{\tau}{0}{t} G_2(t-\tau)[k\lam(\tau)+\eta(\tau)],
\end{equation*}
where the two kernels are given as the inverse Laplace transform of $\hat
G_1(s)=\hat\gam(s)\hat G_2(s)$ and $\hat G_2(s)=[s\hat\gam(s)+k]^{-1}$,
respectively. The system is prepared at time $t=0$ in equilibrium with initial
position $x_0$ drawn from $\peq(x,0)$. Due to the change of the external
parameter $\lam$, the mean
\begin{equation*}
  m(t) \equiv \mean{x(t)} = \IInt{\tau}{0}{t}kG_2(t-\tau)\lam(\tau)
\end{equation*}
is a functional of $\lam(\tau)$. Without loss of generality, we have set
$\lam(0)=0$ and hence $\mean{x_0}=0$.

The substitute operator for one-dimensional Gaussian processes has been worked
out explicitly in~\cite{hang77} reading in general
\begin{equation*}
  \sms(t) = -\partial_x\left[\dot\chi(t)x+\dot\mu(t)
    -\frac{1}{2}\dot\sigma(t)\partial_x\right].
\end{equation*}
The functions $\dot\mu(t)$ and $\dot\sigma(t)$ are determined through the
differential equations
\begin{equation*}
  \dot m(t) = \dot\mu(t) + \dot\chi(t)m(t), \qquad
  \dot v(t) = \dot\sigma(t) + 2\dot\chi(t)v(t)
\end{equation*}
with time-dependent mean $m(t)$ and variance $v(t)$. The correlation function
\begin{equation}
  \label{eq:chi}
  \chi(t,t') \equiv \frac{\mean{[x(t)-m(t)][x(t')-m(t')]}}{v(t')}
\end{equation}
with $\chi(t,t)=1$ determines
$\dot\chi(t)\equiv\partial_\tau\chi(\tau,t)|_{\tau=t}$.

To be more specific, we choose an exponential friction kernel
\begin{equation*}
  \gam(t) = \kap e^{-\kap t} \quad\Rightarrow\quad
  \hat\gam(s) = \frac{\kap}{s+\kap} \quad\Rightarrow\quad
  G_2(t) = (\bar\kap/k)^2 e^{-\bar\kap t} + \frac{\delta(t)}{\kap + k}
\end{equation*}
with inverse time scale $\bar\kap\equiv\kap k/(\kap+k)$. In the Markov limit,
$\kap\rightarrow\infty$ yields $\bar\kap\rightarrow k$ as expected. Using the
explicit expression for the kernel $G_2(t)$, we calculate the mean
\begin{equation*}
  m(\tau) = e^{-\bar\kap(\tau-\tau')}m(\tau') 
  + \lam\left[1-e^{-\bar\kap(\tau-\tau')}\right]
\end{equation*}
where we have stopped the process at $\tau'$ with parameter
$\lam=\lam(\tau')$. This equation shows the basic features of ergodic
non-Markovian processes. For fixed $\lam$, the mean
$m(\tau\rightarrow\infty)\rightarrow\lam$ relaxes towards this value. It is a
functional of $\lam(\tau)$ up to $\tau'$ and afterwards depends on the time
difference $\tau-\tau'$ only. The time derivative yields $\dot
m(\tau)=-\bar\kap m(\tau)+\bar\kap\lam$ and indeed a straightforward
calculation of~(\ref{eq:chi}) confirms $\dot\chi=-\bar\kap$.  Therefore, we
have $\dot\mu=\bar\kap\lam$ and since we do not change the strength of the
trap, the variance is $v=1/(\beta k)$ leading to $\dot\sigma=2\bar\kap/(\beta
k)$. Hence, the substitute operator for fixed $\lam$ becomes
\begin{equation*}
  \sms = \bar\kap\partial_x\left[(x-\lam) + \frac{1}{\beta k}\partial_x\right]
\end{equation*}
with stationary solution $\peq(x,\lam)\propto\exp[-\beta(k/2)(x-\lam)^2]$ for
all times $\tau\geqslant\tau'$.


\section*{References}


\end{document}